# Capturing the "Whole Tale" of Computational Research: Reproducibility in Computing Environments

Bertram Ludäscher, School of Information Sciences, University of Illinois at Urbana-Champaign; Kyle Chard, University of Chicago; Niall Gaffney, Texas Advanced Computing Center, University of Texas at Austin; Matthew B. Jones, University of California Santa Barbara; Jaroslaw Nabrzyski, University of Notre Dame; Victoria Stodden,* School of Information Sciences, University of Illinois at Urbana-Champaign; and Matthew Turk, School of Information Sciences, University of Illinois at Urbana-Champaign
*Corresponding author address: School of Information Sciences, University of Illinois at Urbana-Champaign, Champaign, IL 61820, USA; email: vcs@illinois.edu

*Abstract:* We present an overview of the recently funded "Merging Science and Cyberinfrastructure Pathways: The Whole Tale" project (NSF award #1541450). Our approach has two nested goals: 1) deliver an environment that enables researchers to create a complete narrative of the research process including exposure of the data-to-publication lifecycle, and 2) systematically and persistently link research publications to their associated digital scholarly objects such as the data, code, and workflows. To enable this, Whole Tale will create an environment where researchers can collaborate on data, workspaces, and workflows and then publish them for future adoption or modification. Published data and applications will be consumed either directly by users using the Whole Tale environment or can be integrated into existing or future domain Science Gateways.

## 1. Introduction

Computational resources and scientific services are now nearly ubiquitous in scientific investigations; however, the applications used to discover and analyze data are extremely fragmented and can be intractable, creating a large and meaningful gap between the research processes and the ability to verify the findings [1]. There is frequently no way to trace findings in publications back through the originating computations and data. The Whole Tale project aims to remedy this gap in two ways: 1) integrate existing cyberinfrastructure that supports the entire computational process underlying discovery, thus simplifying the ability for researchers to conduct computational research; and 2) and capture and deliver relevant workflow and processing provenance that will be discoverable and accessible from the associated publication. Whole Tale envisions a collaborative environment where data providers, application developers, and data consumers collaborate and create end-to-end workflows converting data to information using reproducible computational methods.

## 2. The Whole Tale Research Environment

Whole Tale will enable a research environment that seamlessly supports computational tools for tackling pressing research problems in a way that is scalable and reproducible but that still supports software familiar to current researchers. Our aim is to support scientific investigation at all computational scales, from HPC environments to single-user endeavors (the "long tail" of science). We will provide a research environment that captures and, at the time of publication, exposes salient details of the research via access to persistent versions of the data and code used, workflow provenance, data lineage, parameter settings, and output data. *Our approach differs, and is complementary to, that provided by some science gateways in that we rely on utilization of commodity tools, rather than bespoke, domain-specific instruments.*

The Whole Tale environment will provide linkages to existing cyberinfrastructure to provide a research environment that will be instrumented with workflow and reproducibility tools to aid in capturing and storing key scripts, function calls, parameter settings and machine state information that are essential for reproducing the results.

The cyberinfrastructure will be exposed to users through well-known applications such as Jupyter Notebooks that support commonly used data analysis languages including R and Python. Storage will be exposed to users through several interfaces including Globus, a web based filesystem interface, FUSE modules for filesystem-level access to local and remote data repositories, the DataONE federation of data repositories, and an open source Cloud storage environment Nextcloud. By building data repository access into modules that present file-like interfaces, we further lower the barrier to access for remote data stores. The system will also incorporate Globus Auth—a unified an identity management system that will allow users to leverage their own campus, ORCID identifier, or other existing identities. Whole Tale will also enable the deployment of Dockerfile-based environments to support extensible and customizable research workflows.

A lesson we have learned, from this and other cyberinfrastructure projects, is that close collaboration with the research community is essential to success in development and in uptake. To this end we have initiated community driven topical working groups to pilot and provide feedback and use cases from the very beginning of the project.

Novel contributions include the development of authentication systems that seamlessly provide access to external repositories and compute systems, the linking of data resources with compute resources in such a way that enables provenance tools to capture the steps taken in deriving the findings. Additionally, by focusing effort on tools already used across disciplines (such as Jupyter and its ecosystem) we will directly focus on reducing the overall difficulty and complexity of uptake of Whole Tale components in research environments.

## 3. Dissemination of Research Findings

The Whole Tale infrastructure will deliver *research compendia* [2] as research output, comprising not only the publication, but also any data, code, and workflows upon which the findings depend. It is vital that these digital scholarly objects are discoverable, especially to readers of the resulting publication. We propose embedding persistent links to these objects within the publication, Digital Object Identifier assignment, and persistent accessible storage using trusted repositories and other methods. Such practices will enable such objects to be discoverable and citable work.

The reproducibility aspects of the Whole Tale project enable not only the original research to "replay" their discovery pipelines to regenerate their findings, but also for other researchers to do so as well (given appropriate permissions) We have foregrounded reproducibility of computational research and researcher productivity as the principal deliverables we intend to provide.

## 4. Conclusion

The Whole Tale project can be considered a form of generic science gateway: the research environment aims to abstract interactions between researchers and cyberinfrastructure providers. However, its greatest benefits may be leveraged from the collection of microservices (e.g., data access, persistent identifier creation, etc.) and interoperability software that could be built upon and extended by other science gateways. Furthermore, we intend to develop support for community gateways in the research environment to enable researchers to source data and perform analyses from multiple science gateways, while tracking provenance and enabling linkage of processes and data with publications.

## 5. Acknowledgments

Whole Tale is supported by NSF Award #1541450.